# EFFECTS OF DISORDER ON SUPERCONDUCTIVITY
# OF SYSTEMS WITH COEXISTING
# ITINERANT ELECTRONS  AND LOCAL PAIRS

G. Pawłowski, R. Micnas and S. Robaszkiewicz

*Faculty of Physics, A. Mickiewicz University*

*ul. Umultowska 85, 61-614 Poznań, Poland*

## Abstract

We study the influence of diagonal disorder (random site energy) of local pair (LP) site energies on the superconducting properties of a system of coexisting local pairs and itinerant electrons described by the (hard-core) boson-fermion model.

Our analysis shows that the properties of such a model with s-wave pairing can be very strongly affected by the diagonal disorder in LP subsystem (the randomness of the LP site energies). This is in contrast with the conventional s-wave BCS superconductors, which according to the Anderson's theorem are rather insensitive to the diagonal disorder (i.e. to nonmagnetic impurities). It has been found that the disorder effects depend in a crucial way on the total particle concentration $n$ and the LP level position $\Delta_o$ and depending on the parameters the system can exhibit various types of superconducting behaviour, including the LP-like, intermediate (MIXED) and the "BCS"-like.

In the extended range of $\{n, \Delta_o\}$ the superconducting ordering is suppressed by the randomness of the LP site energies and the increasing disorder induces a changeover from the MIXED-like behaviour to the BCS-like one, connected with abrupt reduction of $T_c$ and energy gap to zero. However, there also exist a definite range of $\{n, \Delta_o\}$ in which the increasing disorder has a quite different effect: namely it can substantially enhance $T_c$ or even lead to the phenomenon which can be called disorder induced superconductivity. Another interesting effect is a possibility of a disorder induced bound pair formation of itinerant electrons, connected with the change-over to the LP-like regime.

# 1. Introduction

A model of coexisting bound electron pairs and itinerant electrons i.e. the so-called boson-fermion model, was introduced several years ago [1] to unify the description of nonconventional (exotic) superconductors, chalcogenide glasses, nonsimple metals and systems with alternating valence. Real space local pairing plays an important role in a great number of experimental systems (for review see [1-3] and ref. therein). These systems comprise several distinct groups of materials:

(i) The compounds that contain cations in two valence states differing by 2e (on-site pairing). Examples of such valence skippers are $Bi^{4+}$, $Sb^{4+}$, $Pb^{3+}$, $Sn^{3+}$, $Tl^{2+}$. These elements prefer to exist as $Bi^{5+}$, $Sb^{5+}$, $Pb^{4+}$, $Sn^{4+}$, $Tl^{3+}$ together with $Bi^{3+}$, $Sb^{3+}$, $Pb^{2+}$, $Sn^{2+}$, $Tl^{1+}$. Well studied materials are $BaBi_xPb_{1-x}O_3$, $Ba_{1-x}K_xBiO_3$ ($Bi^{3+}$, $Bi^{5+}$), $Pb_{1-x}Tl_xTe$ ($Tl^{1+}$, $Tl^{3+}$), $Pb_{1-x}Sn_xSe$ ($Sn^{2+}$, $Sn^{4+}$), $Cs_2SbCl_3$ ($Sb^{3+}$, $Sb^{5+}$) as well as divalent compounds of $Ag^{2+}$, $Au^{2+}$ and trivalent compounds of $Pd^{3+}$, $Pl^{3+}$, e.g. $CsAgCl_3$ ($d_{z^2}^2$, $d_{z^2}^0$). Double charge fluctuations occur in these systems on effective sites which can be considered as negative U centers, composed of cations and their surrounding ligand environments.

(ii) The transition metal oxides showing intersite bipolarons, in which charge dispropotionation is seen on a molecular (rather than atomic) level, e.g. $Ti_4O_7$, $Ti_{4-x}V_xO_7$ and $WO_{3-x}$ with double charge fluctuations on the molecular units: [($Ti^{4+}$- $Ti^{4+}$), ($Ti^{3+}$- $Ti^{3+}$)], [($W^{5+}$-$W^{5+}$), ($W^{6+}$-$W^{6+}$)].

Among various other systems, where the local electron pairing has been evidenced [1, 2] let us mention such exotic superconductors as $PbH_x$, $Li_{1-x}Ti_{2-x}O_4$, $Eu_xMo_6S_8$, superconducting nonsimple metals, e.g. NbC, as well as numerous amorphous semiconductors.

The scenario of the coexistence of on-site LPs (negative U centers) or intersite LPs and itinerant carriers can serve as phenomenological description of unconventional superconductivity and normal state properties observed in a number of the above listed compounds. The model is capable of explaining several anomalous properties of these materials and it can show features which are intermediate between those of local pair superconductors and those of classical BCS systems [1-8].

Such a two-component model is of relevance for high temperature superconductors (HTS) and other short-coherence-length superconductors [2-21] as well as for the BCS – Bose-Einstein con-densation (BEC) crossover in ultracold fermionic atomic gases with a Feshbach resonance [22-25].

In the context of HTS the two-component boson-fermion (BF) model has been proposed phenomenologically or it has been derived as the effective low-energy model. In particular, it has been obtained from the generalized periodic Anderson model with on-site hybridization between wide and narrow-band electrons, in which the narrow band electrons are strongly coupled with the lattice deformation, and formation of polarons and LP (bipolarons) takes place [1]. These LP are hard-core bosons made up of two tightly bound fermions. The BF scenario has also been proposed in studies of the superconductivity mechanism based on heterogeneity of the electronic structure of HTS in the pseudogap phase [5, 10-12].

The superconducting and normal state characteristics of the B-F model are strongly dependent on the relative position of the local pair level with respect to the bottom of the fermion band ($\Delta_o$) and the total number of carriers $n$. As a consequence, the model exhibits not only the limiting cases of weak coupling ("BCS") and pure LP superconductivity, but also the intersting, intermediate regime (MIXED).

In this case, neither the standard BCS picture nor the picture of local pairs fits, and superconductivity has a "mixed" character with a correlation length of the order of several interatomic spacings. The system shows features which are intermediate between the BCS and preformed local pair regimes. This concerns the energy gap in the single electron spectrum, $k_B T_C / E_g(0)$ ratio, the critical magnetic fields, the Ginzburg ratio $\kappa$, the width of the critical regime and the normal state properties, which deviate from Fermi liquid behaviour [1-5,7,12] .

In particular, several studies of the boson-fermion model at $T > T_C$ show: the existence of a pseudogap in single particle DOS, anomalies in one-electron self-energy $\sum(\vec{k},\omega)$, and anomalies in charge and magnetic responses, which are similar to those observed in cuprates [7, 24].

In addition, the Uemura type plots, i.e., the $T_C$ vs zero-temperature phase stiffness $\rho_S(0)$, are obtained for s, extended s- and d-wave symmetry within the KT scenario [12]. The Uemura scaling $T_C \propto \rho_S(0)$ is a consequence of the separation of energy scales for pairing and phase coherence. The existence of nodal quasiparticles for d-wave pairing (beyond the LP regime) gives rise to a linear  in $T$ relationship in the superfluid density.

In this paper we will analyze the effects of disorder on the s-wave superconductivity of the B-F model and focus on the diagonal disorder in LP subsystem which is modelled by the random-ness of LP site energies. Our preliminary studies of this subject have been presented in Refs. [16, 18].

The conventional s-wave symmetry superconductors are known to be rather weakly affected by nomagnetic impurities [26] ("Anderson theorem"). Because of the nonconventional pairing mechanism in the B-F model (the charge exchange between bosons and fermions), it is of fundamental importance to determine how the impurities (disorder) affect the bulk and local properties of the superconducting state of such a system [16-21]. We add that the superconductivity in several above mentioned materials is obtained upon doping of the parent compound. Doping is the source of mobile charge carriers and at the same time it introduces disorder into the system.

An example of real systems, where the model of coexisting local pairs and itinerant carriers is applicable and where the diagonal disorder of local pair site energies can be realized, is the family of doped barium bismuthates. Experimental results indicate an existence of p-holes in doped $BaBiO_3$ perovskites [27, 28]. This suggests that a realistic modelling of these materials should be that of coexisting itinerant holes (on oxygen-ions) and LPs (on Bi-ions) in which doping introduces a disorder of the LPs ($Bi^{3+}$ - $Bi^{5+}$) energies.

In the case of cuprate HTS, an indirect support for the two-component B-F model with disorder might be found in the recent scanning-tunneling microscopy (STM) studies [29-31], showing well defined spatial variations in gap magnitude, which correlate with specific impurity sites in the Bi-based cuprates (BiSrCaCuO crystals). Very recently theoretical studies of the local properties of HTS relevant to the STM experiments have been performed in the framework of the B-F model with disorder, using the Bogoliubov-de Gennes equations [20, 21]. It has been demonstrated that the assumption that impurities mainly modify the boson energies offers a natural explanation of the above mentioned correlation found in the STM experiments [21].

We consider the following model Hamiltonian describing two coexisting subsystems: local electron pairs (hard-core bosons "b") and itinerant (conduction) electrons ("c"), which, in the following, will be called c-electrons

$$H = H_b + H_c + H_{cb},$$ (1)

where

$$H_b = 2\sum_i (\Delta_o + E_i - \mu) n_i^b - \sum_{ij} J_{ij} b_i^+ b_j,$$ (2)

$$H_c = \sum_{ij\sigma} t_{ij} c_{i\sigma}^+ c_{j\sigma} - \mu \sum_i n_i^c,$$ (3)

$$H_{cb} = I_o \sum_i (c_{i\uparrow}^+ c_{i\downarrow}^+ b_i + h.c.) \, , \qquad\qquad (4)$$

$\Delta_o$ measures the relative position of the local pair (LP) level with respect to the bottom of c-electron band, $E_i$ is the random LP site energy, $\mu$ stands for the chemical potential, $I_o$ is the intersubsystem coupling (charge exchange), $J_{ij}$ denote the LP transfer integral, $t_{ij}$ is the hopping integral for c-electrons. The operators of local pairs $b_i^+$, $b_i$ obey the commutation rules for hard-core bosons (the Pauli spin $1/2$ commutation rules), which exclude multiple occupancy of a given LP center:

$$[b_i, b_j^+] = (1 - 2n_i^b)\delta_{ij}, \; [b_i, b_j] = 0, \; (b_i^+)^2 = (b_i)^2 = 0, \; b_i^+ b_i + b_i b_i^+ = 1,$$

where $n_i^b = b_i^+ b_i$ and $n_i^c = \sum_\sigma c_{i\sigma}^+ c_{i\sigma}$ are the number operators for bosons and fermions, $\mu$ is the chemical potential which ensures that a total number of particles per site is constant $n = n_c + 2n_b$, where $n_c$ is the concentration of c-electrons and $n_b$ is the average number of local pairs per site.

The paper is organized as follows.

In Sec. 2 we introduce the basic definitions, point out details of the variational approach used in the analysis and present the selfconsistent equations for the order parameters and chemical potential as well as the free energies of the superconducting and the normal state derived at the mean-field level. Sec. 3 includes numerical calculations of the phase diagrams and superconducting characteristics of the model. The phase diagrams at T = 0 and at finite temperatures are determined as a function of particle concentration and the strength of disorder for several representative values of $\Delta_o$.

The evolution of superconducting characteristics, including order parameters, gap ratio, chemical potential, $n_c$ and $n_b$, with increasing disorder is analysed. Sec. 4 is devoted to conclusions and supplementary discussion.

## 2. General Formulation

In analysis of the model we used, as in the case without disorder [1, 2, 8], the mean field variational approach (MFA-HFA). At first, the variational free energy $F_o(\{E_i\})$ for a given fixed configuration of the random site energy $\{E_i\}$ is obtained. Then it is configurationally averaged over the random variable $\{E_i\}$ according to a preset probability distribution $P(\{E_i\})$ [16, 32, 33] as

$$< ... >_{av} = \int\limits_{-\infty}^{\infty} \prod dE_i P(\{E_i\}) \, ... \, . \tag{5}$$

The probability distribution $P(\{E_i\})$ of $\{E_i\}$ is assumed to be $P(\{E_i\}) = \prod\limits_i p(E_i))$, with $p(E_i) = p(-E_i)$. In our studies we have considered the following two types of the site energy distribution:

*the two-delta distribution*
$$p(E_i) = (1/2)[\delta(E_i - E_o) + \delta(E_i + E_o)] \tag{6}$$

and *the rectangular distribution*

$$p(E_i) = 1/2E_o, \qquad \text{for } |E_i| \leq E_o,$$
$$= 0, \qquad \text{otherwise.} \tag{7}$$

In this paper we will mainly focus on the case of distribution (6) (Sec. 3.1-3.3, 3.5) and the results for distribution (7) will be concluded in Sec. 3.4.

The superconducting state (SC) is characterized by the two local order parameters $< c_{i\uparrow}^+ c_{i\downarrow}^+ >$ and $< b_i^+ >$. After the MFA decoupling the effective Hamiltonian is of the form:

$$H_o = 2\sum_i (\Delta_o + E_i - \mu)n_i^b + \sum_{ij\sigma} t_{ij} c_{i\sigma}^+ c_{j\sigma} - \mu \sum_i n_i^c +$$

$$- \sum_i \left(\Omega_i b_i^+ + h.c.\right) + \frac{1}{2} \sum_{ij} J_{ij} \left(\left\langle b_i^+ \right\rangle \left\langle b_j \right\rangle + h.c.\right) +$$

$$+ I_o \sum_i \left(\left\langle c_{i\uparrow}^+ c_{i\downarrow}^+ \right\rangle b_i + h.c.\right) + I_o \sum_i \left(c_{i\uparrow}^+ c_{i\downarrow}^+ \left\langle b_i \right\rangle + h.c.\right) +$$

$$- I_o \sum_i \left(\left\langle c_{i\uparrow}^+ c_{i\downarrow}^+ \right\rangle \left\langle b_i \right\rangle + h.c.\right), \tag{8}$$

where $\Omega_i = \sum_j J_{ij} \langle b_j \rangle$.

In the following we will assume a uniform and site-independent order parameters:

$$\left\langle \left\langle b_i^+ \right\rangle \right\rangle_{av} = \rho_o^x, \ \rho_o^x = \frac{1}{2N} \sum_i << b_i^+ + b_i >>_{av}, \ \left\langle \left\langle c_{i\uparrow}^+ c_{i\downarrow}^+ \right\rangle \right\rangle_{av} = x_o,$$

$x_o = \frac{1}{N} \sum_i << c_{i\uparrow}^+ c_{i\downarrow}^+ >>_{av}$, leaving the full discussion of the Bogoliubov- de Gennes equations

to a separate paper.

Under the above assumptions the quenched free energy $< F_o(\{E_i\}) >_{av}$ is derived as:

$$< F_o^s >_{av} = -\frac{2}{\beta N} \sum_{\vec{k}} \ln\left[ 2\cosh\left(\frac{\beta A_{\vec{k}}}{2}\right) \right] - \frac{1}{\beta} < \ln[2\cosh(\beta \xi_i)] >_{av} + \mu(2n_b + n_c) +$$

$$- \mu - \overline{\mu} + C, \tag{9}$$

where

$$A_{\vec{k}} = \sqrt{\overline{\varepsilon}_{\vec{k}}^2 + (I_o \rho_o^x)^2}, \ \xi_i = \sqrt{(\overline{\mu} - E_i)^2 + (J_o \rho_o^x - I_o \mid x_o \mid)^2}, \tag{10}$$

$$\overline{\mu} = \mu - \Delta_o, \ \ \overline{\varepsilon}_{\vec{k}} = \varepsilon_{\vec{k}} - \mu, \ \ \ \varepsilon_{\vec{k}} = \widetilde{\varepsilon}_{\vec{k}} - \varepsilon_b = -2t \sum_{(\alpha=x,y,...)} \cos k_\alpha a - \varepsilon_b, \ \ \varepsilon_b = \min \widetilde{\varepsilon}_{\vec{k}},$$

$$J_o = \sum_{j\neq i} J_{ij}, \ \ \beta = (k_B T)^{-1}, \ C = J_o(\rho_o^x)^2 - 2I_o \mid x_o \mid \rho_o^x - \varepsilon_b,$$

$$n_c = \frac{1}{N} \sum_i << n_i^c >>, \ \ n_b = \frac{1}{N} \sum_i << n_i^b >>_{av}, \tag{11}$$

whereas the equations determining the superconducting order parameters $\rho_o^x$, $x_o$ and the chemical potential are given by:

$$x_o = -\frac{1}{N} \sum_{\vec{k}} \frac{I_o \rho_o^x}{2 A_{\vec{k}}} \tanh(\beta A_{\vec{k}} / 2), \tag{12}$$

$$\rho_o^x = \left\langle (-I_o x_o + J_o \rho_o^x) \frac{1}{2\xi_i} \tanh(\beta \xi_i) \right\rangle_{av}, \tag{13}$$

$$n = n_c + 2n_b, \tag{14}$$

where

$$n_c - 1 = -\frac{1}{N}\sum_{\bar{k}}\frac{\bar{\varepsilon}_{\bar{k}}}{A_{\bar{k}}}\tanh\left(\frac{\beta A_{\bar{k}}}{2}\right), \qquad 2n_b - 1 = -\left\langle\frac{\Delta_o + E_i - \mu}{\xi_i}\tanh(\beta\xi_i)\right\rangle_{av}. \qquad (15)$$

The quasiparticle energy spectrum of the system is characterized by $E_k^{\pm} = \pm A_k$ and $\pm\xi_i$, and the energy gap in the electron spectrum is given by $E_g = \min E_k^{+} - \max E_k^{-}$.

From Eqs. (12)-(14) one gets equations for the transition temperature ($T_c^{MFA}$) at which the gap amplitude vanishes ($x_o \to 0, \rho_o^x \to 0$):

$$1 = \left\langle\frac{1}{2(\Delta_i - \mu)}\tanh\left[\beta_c^{MFA}(\Delta_i - \mu)\right]\right\rangle_{av}\left(\frac{I_o^2}{N}\sum_{\bar{k}}\frac{\tanh(\beta_c^{MFA}\bar{\varepsilon}_{\bar{k}}/2)}{2\bar{\varepsilon}_{\bar{k}}} + J_o\right), \qquad (16)$$

$$n = 2 - \left\langle\tanh\left[\beta_c^{MFA}(\Delta_i - \mu)\right]\right\rangle_{av} - \frac{1}{N}\sum_{\bar{k}}\tanh(\beta_c^{MFA}\bar{\varepsilon}_{\bar{k}}/2), \qquad (17)$$

where $\Delta_i = \Delta_o + E_i$. Note that in Eq. (16) the factor $I_o^2/2(\Delta_i - \mu)$ can be interpreted as the pairing interaction between c-electrons mediated by the LP and the factor $\tanh[\beta_c^{MFA}(\Delta_i - \mu)]$ results from the hard-core nature of bosons.

It is also of interest to investigate the response function in the absence of disorder, i.e. $E_i = 0$. Let us consider the pair propagator for c-electron subsystem

$$G_2(\bar{q}, \omega) = \frac{1}{N}\left\langle\left\langle B_{\bar{q}} | B_{\bar{q}}^{+}\right\rangle\right\rangle_{\omega}, \qquad (18)$$

where $\left\langle\left\langle B_{\bar{q}} | B_{\bar{q}}^{+}\right\rangle\right\rangle_{\omega}$ is the time Fourier transform of the Green's function: $-i\Theta(t - t') < B_{\bar{q}}(t), B_{\bar{q}}^{+}(t') >$, $B_{\bar{q}} = \sum_{\bar{k}}c_{-\bar{k}+\bar{q}/2\downarrow}c_{\bar{k}+\bar{q}/2\uparrow}$. In the normal state, using the equation of motion technique and Random Phase Approximation (RPA) for the case $J_{ij} = 0$, one gets:

$$G_2(\bar{q}, \omega) = \frac{\chi(\bar{q}, \omega)}{1 - V_{eff}(\omega, \beta)\chi(\bar{q}, \omega)}, \qquad (19)$$

where

$$\chi(\vec{q},\omega) = \frac{1}{N} \sum_{\vec{k}} \frac{1 - f(\bar{\varepsilon}_{\vec{k}+\vec{q}/2}) - f(\bar{\varepsilon}_{\vec{k}-\vec{q}/2})}{\omega - \bar{\varepsilon}_{\vec{k}+\vec{q}/2} - \bar{\varepsilon}_{\vec{k}-\vec{q}/2}}, \tag{20}$$

$$V_{eff}(\omega,\beta) = \frac{I_o^2}{\omega - 2(\Delta_o - \mu)}(1 - 2n_b), \tag{21}$$

$-\chi(\vec{q},\omega)$ is the pair susceptibility and $V_{eff}(\omega)$ is the effective interaction between c-electrons mediated by LP's. $f(x)$ is the Fermi function. Introducing the generalized T-matrix $\Gamma(\vec{q},\omega)$ via the relation $G_2 = \chi\Gamma/V_{eff}$, one has

$$\Gamma(\vec{q},\omega) = \frac{V_{eff}(\omega)}{1 - V_{eff}(\omega)\chi(\vec{q},\omega)}. \tag{22}$$

An instability of the normal phase occurs at $\Gamma^{-1}(0,0) = 0$ (the Thouless criterion), and $T_c$ is given by

$$1 = -V_{eff}(0,\beta_c^{MFA}) \frac{1}{N} \sum_{\vec{k}} \frac{\tanh(\beta_c^{MFA}\bar{\varepsilon}_{\vec{k}}/2)}{2\bar{\varepsilon}_{\vec{k}}}, \tag{23}$$

in full agreement with the MFA expression (16) for a clean system if we take $1 - 2n_b = \tanh\left[\beta_c^{MFA}(\Delta_o - \mu)\right]$.

The randomness of LP energies $\Delta_i$ induces fluctuations of the effective pairing potential between electrons $V_{eff}$ which we can replace by the following one at $T_c$:

$$V_{eff}(0,\beta) = \left\langle \frac{-I_o^2}{2(\Delta_O - \mu)}(1 - 2n_i^b) \right\rangle_{av}, \tag{24}$$

and $1 - 2n_i^b$ is given by $\tanh\left[\beta(\Delta_i - \mu)\right]$.

The MFA transition temperature ($T_c^{MFA}$) at which the gap amplitude vanishes, yields an estimation of the c-electron pair formation temperature. Generally, due to the fluctuation effects the superconducting phase transition occurs at a critical temperature lower than that predicted by the BCS-MFA theory [12].

## 3. Results of numerical solutions and discussion

We have performed a detailed analysis of the phase diagrams and superconducting properties of the system studied as a function of $\Delta_o$, the particle concentration $n$, the interaction parameters and increasing disorder. For c-electrons we have used a semielliptic density of states (DOS): $\rho(\varepsilon) = (2/\pi D)\sqrt{1 - [(\varepsilon - D)/D]^2}$, with $2D$ denoting a bandwidth and the rectangular one: $\rho(\varepsilon) = 1/2D$, for $0 \le \varepsilon \le 2D$, $\rho(\varepsilon) = 0$, otherwise.

The semielliptic DOS approximates the behaviour of $d = 3$ simple cubic lattice, yielding for small energies: $\rho(\varepsilon) \approx \dfrac{2\sqrt{2}}{\pi D^{3/2}}\sqrt{\varepsilon}$. The rectangular one mimics better the $d = 2$ square lattice spectrum beyond the Van-Hove singularity.

In Sec. 3.1-3.3, 3.5 for the random LP site energies we have taken the two-delta distribution (6). The results for the rectangular distribution of $E_i$ (Eq.(7)) are summarized in Sec. 3.4. We will discuss first the case of $J_0 = 0$ (Sec. 3.1 –3.4) and the case of $J_0 \ne 0$ will be concluded in Sec. 3.5.

### 3.1. Ground state diagrams

Fig. 1 presents the ground state diagram of the model (1) for $\mid I_o \mid = 0$, $J_o = 0$, $E_i = 0$ as a function of $n = n_c + 2n_b$ versus $\Delta_o / D$ $(D = zt)$ plotted for *a square lattice (thick solid lines) semieliptic DOS (thin solid line)* and *rectangular DOS (dashed lines)*.

In the absence of interactions and $E_i = 0$ depending on the relative concentration of electrons and hard-core bosons we distinguish three essentially different physical situations. In particular for $n \le 2$ they are:

(i) $\Delta_o < 0$ so that at $T = 0K$, $\mu = \Delta_o < 0$ and all the available particles occupy the LP states (it will be called the "local pair" regime, in which in general $2n_b >> n_c$ ) (LP);

(ii) $\Delta_o > 0$ such that the electron band is filled up to the Fermi level $\mu = \Delta_o$ and the remaining particles are in local pair states (the c+b regime or MIXED, $0 < 2n_b$, $n_c < 2$ ) (LP+E);

(iii) $\Delta_o > 0$ such that the Fermi level $\mu < \Delta_o$ and consequently at $T = 0K$ all available electrons occupy the c-electron states (the c-regime, $n_c >> 2n_b$) (E).

For $E_i = 0$ an important feature of the model is that the boson-fermion coupling $|I_o| \neq 0$ induces superconducting order in both subsystems at low temperatures. A small $|I_o|$ does not change much the characteristic lines of the diagrams in Fig. 1 concerning the particle densities. However, it yields $n_c \neq 0$ in the superconducting LP regime and $n_b \neq 0$ in the superconducting E regime, even at $T = 0$. Moreover, it also leads to a renormalization of the bosonic level and a shift of the MIXED-LP boundary. The increasing $|I_o|$ expands the range of the LP+E (MIXED) regime [1, 12].

In Figs. 2 we show the ground state diagrams as a function of $n$ and $E_o / 2D$ for $|I_o| = 0$ and $|I_o| / 2D = 0.1$, plotted for several representative fixed values of $\Delta_o / D$ and $|J_o| = 0$.

In the presence of $E_i$ disorder with the two-delta distribution (6) in addition to the former three situations, the fourth one appears possible, in which the occupations of both subsystems are pinned at (almost) fixed values: $n_c = n - 1 < 2$ and $2n_b = 1$ in extended ranges of total concentration n and temperature.

The fourth possibility occurs if the bosonic level with the energy $\Delta_o - E_o$ is completely occupied ($2n_b = 1$) since it lies below the Fermi level ($\mu > \Delta_o - E_o$), whereas the level with the energy $\Delta_o + E_o$ is empty ($\mu < \Delta_o + E_o$). The properties of this state are similar to those of the E state ("BCS" like regime) in the absence of disorder, and therefore we will call it in the following as the E1 state.

Let us point out that, the states of this type will appear in definite ranges of $\Delta_o / D$ and n for any multiple-pole distribution of $E_i$ disorder.

At $T = 0$ and $|I_o| \neq 0$ the boundary between MIXED and LP regimes can be located (after Legget [34]) from the condition that the chemical potential in the SC phase ($\mu_s$) reaches either the bottom (if $n < 2$) or the top of the fermionic band (if $n > 2$) i.e., in our denotations from $\mu_s = 0$, if $n < 2$, or $\mu_s = 2D$, if $n > 2$.

The approximate boundaries MIXED/E and MIXED/E1 for $|I_o|\neq 0$ are plotted in Figs 2 by dashed lines. These lines demark the values of the $n$ and $E_o/2D$ for which $T_c$ and $\rho_o^x$, $x_o$ at $T=0$ become vanishingly small and for which at $T=0$ $2n_b/n$ (or $(2-2n_b)/n$) $<10^{-6}$ – in the case of MIXED/E and $|2n_b-1|/n<10^{-6}$ – in the case of MIXED/E1.

The analysis of the MIXED-LP crossover indicates that when the LP level is shifted and reaches either the bottom of the fermionic band (if $n<2$) or the top of the band (if $n>2$), the effective attraction between fermions $V_{eff}(0)$ (Eq. (24)) becomes strong, since it varies as $\left\langle I_o^2/(2\Delta_i-2\mu)\right\rangle_{av}$ and locally $\mu\approx\Delta_i$. In this regime on the LP side the density of $c$-electrons (if $n<2$) or $c$-holes (if $n>2$) is low and formation of bound $c$-electron (or hole) pairs occurs. It gives rise to an energy gap in the single-electron spectrum persisting even in the normal state. In such a case, the superconducting state can be formed by two types of coexisting (hybridized) bosons: preformed $c$-electron (-hole) pairs and LP's [12].

Note that at the MIXED-LP crossover, the chemical potential changes the sign ($n<2$) (cf. Figs. 4a, b). For considered s-wave pairing the energy gap changes from $\frac{1}{2}E_g(0)=|I_o\rho_o^x|$ ($\mu_S>0$) to $\frac{1}{2}E_g(0)=\sqrt{\mu_S^2+|I_o\rho_O^x|^2}$ ($\mu_S<0$). For the latter case and $n_C<<1$, $\mu_S$ can be related to the electron pair binding energy in the two body problem [12].

For d-wave pairing in the absence of disorder one finds that at the point of crossover there is a change from nodal to nodeless behaviour [12]. It remains to be seen how the presence of non-magnetic disorder will modify this crossover.

In the model considered a strong dependence of the superconducting properties on the randomness in bosonic subsystem is a combined effect of the fluctuations of the pairing strength and the changes in the relative occupation of fermionic and bosonic states induced by disorder.

## 3.2. Evolution of superconducting properties with increasing disorder

In Fig. 3 we show the finite temperature phase diagram of model (1) sketched in 3D projection ($k_BT_c/2D$ vs $n$ vs $E_o/2D$) for a fixed $\Delta_o/D$ and $|I_o|/2D$.

The disorder effects depend in an essential way on the total concentration of carriers $n$ and the position of the LP level $\Delta_o$. In general, as for the evolution of the SC properties with increasing disorder, there are different possible types of change-over. It is shown in Figs 4, which present the plots of the critical temperature and several other superconducting characteristics of the model (1.1) as a function of increasing disorder $E_o/2D$ for $\Delta_o/D = 0.5$, $|I_o|/2D = 0.1$ and for five representative values of n. In all cases the plotted quantities are (from the top): (1) the MFA critical temperature $k_B T_c/2D$ and the chemical potential $\mu/2D$ at $T = 0$, (2) the concentrations of electrons $n_c$ and local pairs $n_b$ at $T = 0$, (3) the superconducting order parameters $\rho_o^x$ and $|x_o|$ at $T = 0$, (4) the gap ratio for c-electrons $E_g(0)/2k_B T_c$.

In Figs. 4. the vertical dotted lines mark the boundaries between various regimes. As we see from these figures and Fig. 2, for $\Delta_o/D = 0.5$ one can single out five types of change-over, each of them being characteristic for a definite range of $n$:

(a) $E \rightarrow MIXED \rightarrow LP$, if $n < 0.5$;

(b) $MIXED \rightarrow LP$, if $0.5 < n < 1$;

(c) $MIXED \rightarrow E1$, if $1 < n < 2.5$;

(d) $E \rightarrow MIXED \rightarrow E1$, if $2.5 < n < 3$;

(e) $E \rightarrow MIXED \rightarrow LP$, if $n > 3$.

These types of change-over also take place for the other fixed values of $\Delta_o/D > 0$ ($0 < \Delta_o/D < 2$). For $\Delta_o/D < 0$ and $1 < n < 2$ an additional sequence of change-over is possible: $LP \rightarrow MIXED \rightarrow E1$ (cf. Fig. 2a).

Let us notice that the change-overs $MIXED \leftrightarrow E$ ($E1$) are quite sharp (cf. Figs. 4c, d). This is connected with an abrupt redistribution of $n_c$ and $n_b$, which take nearly constant values as the system enters the $E(E1)$ regime, and with a rapid decrease in $V_{eff}$ (cf. Fig. 5). The evolution of $T_c$, $E_g(0)$, $n_c/n_b$ and order parameters between the $MIXED$ and $LP$ regimes is smooth (cf. Figs. 4a, b, e). The $MIXED/LP$ boundary is located on the $n$ vs $E_o/2D$ (and $n$ vs $\Delta_o/2D$) diagrams from the condition $\mu_s = 0$, if $n < 2D$ and from $\mu_s = 2D$, if $n > 2$.

The range of $n$ for which a definite sequence of change-overs takes place, depends on the value of $\Delta_o/D$, but in general (in the case of the $E_i$ distribution (6) and $0 < \Delta_o/D < 2$) the $MIXED \leftrightarrow LP$ change-over is possible for $n \leq 1$ and $n \geq 3$ only, whereas the $MIXED \leftrightarrow E1$ - for $1 < n < 3$ (comp. Figs. 2a, b, c).

Let us point out the most striking features of the system considered.

(1) In the extended range of parameters $\{n, \Delta_o / D\}$ even a relatively small disorder can have a strongly detrimental effect on s-wave SC.

In particular, at $\Delta_o / D = 0.5$ it will occur for $1 < n < 2.5$ (comp. Fig. 2b and Fig. 4c), whereas at $\Delta_o / D = 1$ - for $1 < n < 3$ (comp. Fig. 2c), i.e. for the concentrations at which the change-overs $MIXED \leftrightarrow E1$ are realized. In these cases, with increasing disorder the critical temperature $T_c$, the superconducting order parameters $\rho_o^x$, $x_o$ and the energy gap in c-electron spectrum $E_g(0)$ are very strongly suppressed, the gap to $T_c$ ratio is shifted towards the BCS value, whereas $2n_b$ tends to unity (comp. Fig. 4c).

(2) There also exist a range of $\{n, \Delta_o / D\}$ for which the system exhibits a complete opposite behaviour, which can be called a *disorder induced superconductivity*.

At $\Delta_o / D = 0.5$ it can be observed for $0 < n < 0.5$ as well as for $n > 2.5$ (comp. Figs. 2b, 3 and Figs. 4a, d, e), i.e. for the concentrations at which the changeovers $E \rightarrow MIXED$ are realized. In these cases, for $E_o = 0$ the system is in $E$ regime with $T_c \approx 0$. With increasing disorder, at some definite nonzero value of $E_o / 2D$, dependent on $\{n, \Delta_o / D\}$ one observes an abrupt enhancement of $T_c$, the superconducting order parameters and $E_g(0)$, connected with the $E \rightarrow MIXED$ changeover. With a further increase in $E_o / 2D$, $T_c$ goes through a round maximum and then it decreases, either sharply (for the sequence of transitions: $E \rightarrow MIXED \rightarrow E1$), cf. e.g. Fig. 4d, or steadily (for the sequence: $E \rightarrow MIXED \rightarrow LP$), cf. e.g. Fig . 4a.

(3) Another unique feature of the model analyzed is a possibility of *a disorder induced bound pair formation* of itinerant electrons. Such phenomena can be observed for the ranges of $\{n, \Delta_o / D\}$, in which the increasing disorder yields either the following sequence of change-overs: $E \rightarrow MIXED \rightarrow LP$ or a single change - over: $MIXED \rightarrow LP$. In both these cases the bound pair formation occurs close to the $MIXED / LP$ boundary (c.f. e.g. Figs. 4a, 4e), at which the increasing $E_o$ shifts the chemical potential either below the bottom of the c-electron band (if $n \leq 1$) or above the top of the band (if $n \geq 3$).

In Fig. 5 we plot the effective pairing potential between c-electrons $V_{eff}(0, T_c)$ in various regimes of parameters. We see that $V_{eff}$ increases rapidly close to the $MIXED/LP$ boundaries (comp. Figs. 4a, b, e), which induces the bound pair formation of c-electrons (or holes). Inside the $E$ and $E1$ regimes the $V_{eff}$ is small and the properties are unaffected by disorder as in the conventional s-wave BCS superconductors in the presence of nonmagnetic disorder.

### 3.3. Density driven changeovers

Concerning the evolution of the superconducting properties with increasing $n$ there are three possible types of density driven changeovers for $E_o = 0$: (i) for $2 > \Delta_o/D > 0$: $E \rightarrow MIXED \rightarrow E$; (ii) for $\Delta_o/D < 0$: $LP \rightarrow E$; (iii) for $\Delta_o/D > 2$: $E \rightarrow LP$ (comp. Fig. 1).
In the presence of disorder one finds that the system can also exhibit several other sequences of changeovers. In particular, in the case (i) as the disorder amplitude $E_O$ increases the sequence $E \rightarrow MIXED \rightarrow E$ is changed at first into: $E \rightarrow MIXED \rightarrow E1 \rightarrow MIXED \rightarrow E$, then into: $LP \rightarrow E1 \rightarrow MIXED \rightarrow E$ and finally into: $LP \rightarrow E1 \rightarrow LP$ (comp. Figs. 2b, c), whereas in the case (ii) the changeover $LP \rightarrow E$ is replaced by: $LP \rightarrow MIXED \rightarrow E$, then by: $LP \rightarrow E1 \rightarrow MIXED \rightarrow E$ and finally by: $LP \rightarrow E1 \rightarrow LP$ (comp. Fig. 2a).

In Figs. 6 we show examples of the $T_c$ vs $n$ plots, computed for several fixed values of the disorder amplitude $E_o/2D$ and fixed $\Delta_o/D$, which illustrate the change-overs taking place in the case (i).

### 3.4. The case of rectangular distribution of the random potential

In Sec. 3.1-3.3 we focused on random LP energy distribution in a bimodal form (6). As it was demonstrated in our previous studies of the model of hard-core charged bosons on a lattice [32, 33] the detailed features of the phase diagrams and thermodynamic properties can be sensitive to the choice of the distribution function of $E_i$, although the main conclusions concerning profound effects of diagonal disorder remain unchanged.

Here, we will show that the same situation holds as far as the effects of disorder in the boson-fermion model are concerned. To prove this point we present below in Figs. 8-10 the results obtained for the rectangular distribution of $E_i$.

A preliminary analysis of this case has been given in Ref. [18]. Indeed, as in the case of bimodal distribution the disorder effects depend in an essential way on the total concentration of carriers $n$ and the position of the LP level $\Delta_o$. For the rectangular distribution of disorder one can single out four types of change-over characteristic for definite ranges of $n$. With increasing disorder strength $E_o$, the possible change-overs for a fixed value of $\Delta_o > 0$ are (Figs. 7-9):

(i) $E \rightarrow MIXED \rightarrow LP$, (ii) $MIXED \rightarrow LP$, (iii) $E \rightarrow MIXED$, whereas for $\Delta_o < 0$ one additionally finds: (iv) $E \rightarrow LP$.

The qualitative difference between the cases of rectangular and bimodal distribution of $E_i$ is the absence of $E1$ state for the former. As a consequence there are no change-overs involving the $E1$ state and the system can stay in the mixed state (compare Figs. 8 and 3). Furthermore, for both types of $E_i$ distribution, in an extended range of parameters $\{n, \Delta_o\}$ a relatively small disorder has strongly detrimental effect on superconductivity in the mixed regime (cf. Figs. 4 b-c and Fig. 9b). In both cases, a definite range of parameters $\{n, \Delta_o\}$ can be also found for which the boson-fermion model exhibits a disorder induced superconductivity. For the rectangular distribution this phenomenon takes place at around $E \rightarrow MIXED$ change-over (cf. Fig. 7 and Figs. 9a,c).

## 3.5. Disorder effects in the presence of direct LP hopping $J_{ij} \neq 0$

In general, the direct LP hopping $J_{ij} \neq 0$ expands the stability regions of the superconducting LP and MIXED states with respect to the ones of E and E1, and increases the superconducting critical temperatures as well as the superfluid density in both these states.

We postpone a more detailed analysis of the $J_{ij} \neq 0$ case, including the $n$ vs $\Delta_o / D$ vs $E_o / 2D$ phase diagrams, to a separate paper. Here we will focus only on the properties of the system for $J_o \neq 0$, $I_o = 0$ in the limit of $\Delta_o / D << -1$, where $2n_b \rightarrow n$, $n_c \rightarrow 0$, and where the superconductivity can develop exclusively in the LP subsystem (for $I_o = 0$ the c-electrons remain in the normal state and act only as a reservoir of particles).

In Figs. 10-12 we present representative results concerning such a case showing the effects of disorder on the SC phase of LPs for various values of $n_b$ for the two-delta distribution of $E_i$. In particular one finds the following:

(i) With increasing disorder the superconducting order parameter $\rho_o^x$ at $T = 0$ as well as the critical temperatures $T_c$ are strongly reduced for any local pair concentration ($0 < 2n_b < 2$) (cf. Figs. 11, 12).

(ii) For any $n_b$ there is a critical amount of disorder, below which SC can be stable. The critical disorder is the largest close to half filling of the LP band ($2n_b = 1$) and it strongly diminishes with increasing $|2n_b - 1|$ (Figs. 10, 12).

(iii) For $2n_b \neq 1$ the SC transition can be either of second order or first order depending on the local pair concentration $n_b$ and the strength of disorder $E_o$ (cf. Fig. 12). Increasing disorder for $2n_b \neq 1$ changes at first the nature of the phase transition from a continuous to discontinuous type, resulting in the tricritical point (TCP) apperance, then it suppresses the superconductivity at low $n_b$. Finally, for large $E_o / D$ the system remains in the normal state at all $T$ and any $n_b$.

As pointed out in Sec. 3.4 the detailed features of the phase diagrams can be sensitive to the choice of the distribution function of random potential. In particular for $I_o = 0$, $J_o \neq 0$, for *the rectangular distribution* the SC transition is of second order for any local pair concentration $n_b$ and strength of disorder $E_o$ [32, 33].

In the limit $\Delta_o / D << -1$ and $I_o = 0$ the model considered reduces to the model of hard-core bosons on a lattice, which has been studied by Monte-Carlo simulations [35-37] and exact diagonalization of small systems [37-39]. The results of these works, concerning the evolution of critical disorder with particle concentration as well the evolution of superfluid density with $n_b$ for different degrees of disorder and showing that the critical disorder is maximal at $2n_b = 1$ are in good agreement with our findings for this particular case.

## 4. Summary and outlook

In   this paper we have studied the effects of diagonal disorder on the properties of the system of coexisting local pairs and itinerant electrons coupled via charge exchange mechanism, described by the (hard-core) boson-fermion model with random LP site energies. In the analysis we have used the mean field variational approach with a configurationally averaged free energy which fully takes into account the hard core nature of bosons and which in the absence of interactions ($J_{ij} = 0$, $I_o = 0$) yields rigorous results for arbitrary disorder strength. We determined the phase diagrams and superconducting characteristics of the system as a function of the strength of disorder $E_o$ and the total particle concentration $n$, for several representative values of the local pair level position $\Delta_o / D$. Depending on the parameters the model is found to exhibit various types of superconducting behaviour ranging from the "BCS"-like to the local pair-like limits.

Our analysis shows that the properties of the boson-fermion model can be strongly affected by the diagonal disorder in bosonic subsystem (the randomness of the LP site energies). This is in obvious contrast with the conventional s-wave BCS-type superconductors, which according to the Anderson's theorem are rather insensitive to the diagonal disorder (i.e. to nonmagnetic impurities). The remarkable dependence of the superconducting properties on the $E_i$ randomness is a combined effect of changes in the effective pairing interaction and those in the occupation of fermionic and LP states induced by disorder. It has been found that in the model considered, the effects of diagonal disorder depend in a crucial way on the particle concentration and the LP level position $\Delta_o$. In the extended range of $\{n, \Delta_o\}$ the superconducting ordering can be strongly suppressed by the randomness of the site energies (cf. Figs. 4b, c, d, Fig. 9b) and the increasing disorder can induce a change-over $MIXED \rightarrow E1$ (cf. Figs. 3, 4c, d). However, there also exist a definite range of the parameters $\{n, \Delta_o\}$ for which the increasing disorder has a quite different effect: namely it can substantially enhance $T_c$ or even yield the phenomena which can be called *a disorder induced superconductivity* (connected with the changeover $E \rightarrow MIXED$ – cf. Figs. 3, 4a, d, e and Figs. 9a, c). Another interesting effect is a possibility of *a disorder induced bound pair formation of c-electrons*, connected with the change-over $MIXED \rightarrow LP$, at which the effective interaction ($V_{eff}$) between electrons mediated by LPs is found to be very strongly enhanced in comparison to its values inside the MIXED, E and E1 regimes (cf. Fig. 5).

As we found, the detailed features of the phase diagrams can be sensitive to the choice of the distribution function of random potential. Nevertheless, our analysis of two different types of the $E_i$ distribution functions indicates that the basic conclusions concerning profound effects of diagonal disorder on superconductivity in boson-fermion model remain unchanged for both cases.

In our work we have not considered the randomness in the fermion site energies, since weak disorder of this type is responsible mainly for renormalization of the single particle DOS in calculations of $T_c$ [17]. Our mean-field treatment of the disordered boson-fermion model misses important effects related to localization as well as phase fluctuations. They are of special relevance in the LP regime, where the mean-field $T_c$ serves only as a pairing scale, rather than the true transition temperature [1, 12].

In particular, for the nonrandom case, the $T_c s$ as calculated beyond MFA within the T-matrix approach in 3D [12, 24, 25] and in the KT scenario for 2D [12], show the crucial effects of pair fluctuations (and phase fluctuations) in the mixed and LP regimes. Typically and independently of the pairing symmetry, it is observed that in 2D the $T_c^{KT}$ is substantially smaller than $T_c^{MFA}$ in these cases. Similarly, $T_c$ in 3D is strongly reduced as compared to $T_c^{MFA}$ due to pairing fluctuations. In the mixed regime, for temperatures between $T_c^{MFA}$ and $T_c^{KT}$ (or $T_c$ within T-matrix), the system exhibits a pseudogap in the fermionic spectrum, which develops into a real gap when moving to the LP regime. For $J_{ij} = 0$, and beyond MFA, the reduction of $T_c$ in the LP regime is caused by increasing the effective mass of LP's and lowering their mobility. Introducing disorder can yield localization in such a case.

An extension of the present study in the above directions is planned.

## 5. Acknowledgments

We would like to thank K.Wysokinski and T.Domański for useful comments and discussion.

**Figure captions**

**Fig. 1.** Ground state diagram of the model (1) as a function of $n = n_c + 2n_b$ versus $\Delta_o / D$ ($D = zt$) for $|I_o| = 0$, $|J_o| = 0$, $E_o = 0$, plotted for a square lattice (thick solid lines) semielliptic DOS (thin lines) and rectangular DOS (dashed lines). Denotations: LP – nonmetallic state of LP's ($2n_b = n$, $n_c = 0$ for $n < 2$, $2n_b = 2 - n$, $n_c = 2$ for $n > 2$), E – metallic state of electrons ($n_c = n$, $2n_b = 0$ for $n < 2$, $n_c = 2 - n$, $2n_b = 2$ for $n > 2$), LP+E (MIXED) – "mixed" regime ($0 < n_c, 2n_b < 2$).

**Fig. 2.** Ground state diagrams of the model (1) as a function of $n = n_c + 2n_b$ and $E_o / 2D$ ($D = zt$) for $|I_o| = 0$, $J_o = 0$ (solid lines) and $|I_o| / 2D = 0.1$ ($J_o = 0$) (dashed lines), plotted (from the top) for $\Delta_o / D = -0.2$, $0.5$, $1.0$ (a, b, c, respectively, semielliptic DOS and the two-delta distribution of $E_i$ (Eq. 6)). Denotations:

LP – "local pair regime" ($2n_b = n$, $n_c = 0$ for $n < 2$, $2n_b = 2 - n$, $n_c = 2$ for $n > 2$),

E – a "c-electrons" regime ("BCS") ($n_c = n$, $2n_b = 0$ for $n < 2$, $n_c = 2 - n$, $2n_b = 2$ for $n > 2$),

E1 – a second "c-electrons" regime ($n_c = n - 1$, $2n_b = 1$), possible only in the presence of disorder ($E_o > 0$). It occurs if the bosonic level with the energy $\Delta_o - E_o$ is completely occupied ($2n_b = 1$) since it lies below the Fermi level ($\mu < \Delta_o - E_o$), whereas the level with the energy $\Delta_o + E_o$ is empty ($\mu < \Delta_o + E_o$),

MIXED – "mixed" regime ($0 < n_c, 2n_b < 2$).

For $|I_o| \neq 0$, superconductivity sets in all the regimes albeit in E and E1 regions the order parameters are exponentially small.

**Fig. 3.** Finite temperature phase diagrams of the model (1) sketched in three dimensional projection ($k_B T_c / 2D$ vs $n$ vs $E_o / 2D$) for $\Delta_o / D = 0.5$, $|I_o| / 2D = 0.1$ ($J_o = 0$). Dotted lines denote the MIXED/LP boundaries at $T = 0$.

**Fig. 4.** Evolution of the critical temperature and various other superconducting characteristics of the model (1) as a function of increasing disorder $E_o / 2D$ for $\Delta_o / D = 0.5$, $|I_o| / 2D = 0.1$ plotted for several fixed values of $n$: (a) $n = 0.2$, (b) $n = 0.9$, (c) $n = 1.8$, (d) $n = 2.8$, (e) $n = 3.5$ (semielliptic DOS, $J_o = 0$).

In all cases the plotted quantities are (from the top): (i) the critical temperature $k_B T_c / 2D$ (solid line) and the chemical potential $\mu / 2D$ at $T=0$ (dashed line), (ii) the concentrations of electrons $n_c$ (dashed) and local pairs $n_b$ (solid) at $T=0$, (iii) superconducting order parameters $\rho_o^x$ (solid) and $|x_o|$ (dashed) at $T=0$, (iv) the gap ratio for c-electrons $E_g(0) / 2k_B T_c$.

**Fig. 5.** The averaged pairing potential between c-electrons $V_{eff} = |V_{eff}(0, T)|$ at $T_C$ as a function of increasing disorder $E_o / 2D$ for $\Delta_o / D = 0.5$, $|I_o| / 2D = 0.1$ ($J_o = 0$) plotted for several representative values of *n= 0.2, 0.9, 1.8, 2.8, 3.5.*

**Fig. 6.** Phase diagrams in the (T-n) plane for $\Delta_o / D = 0.5$, $|I_o| / 2D = 0.1$ plotted for $E_o / 2D = 0.0$, $E_o / 2D = 0.01$, $E_o / 2D = 0.05$, $E_o / 2D = 0.30$, $E_o / 2D = 0.60$, $E_o / 2D = 0.80$ (semielliptic DOS, $J_o = 0$).

**Fig. 7.** Ground state diagrams of the model (1) for the rectangular distribution of $E_i$ (Eq.(7)) and for rectangular DOS of electrons as a function of $n = n_c + 2n_B$ and $E_o / 2D$ ($D = zt = 4t$) for $|I_o| / 2D = 0.1$ ($J_o = 0$) plotted (from the top) for $\Delta_o / D = 0.1, 0.5$. Denotations as in Fig. 2.

**Fig. 8.** Finite temperature phase diagrams of the model (1) for the rectangular distribution of $E_i$ (Eq.(7)) and for rectangular DOS of electrons sketched in three dimensional projection ($k_B T_C / 2D$ vs n vs $E_O / 2D$) for $\Delta_o / D = 0.5$, $|I_O| / 2D = 0.1$ ($J_o = 0$).

**Fig. 9.** Evolution of the critical temperature $k_B T_c / 2D$ (solid line) and the chemical potential $\mu / 2D$ at T=0 (dashed line) of the model (1) for the rectangular distribution of $E_i$ (Eq.(7)) and for rectangular DOS of electrons as a function of increasing disorder $E_o/2D$ for $\Delta_o/D = 0.5$, $|I_o|/2D = 0.1$ plotted for several values of n: (a) $n = 0.2$, (b) $n = 1.6$, (c) $n = 2.7$.

**Fig. 10.** Ground state diagram of the model (1) as a function of $2n_b$ versus $E_o/2D$ for $J_o/2D = 0.05$, $|I_o|/2D = 0$, $n_b = n/2$.

**Fig. 11.** The superconducting order parameter $\rho_o^x$ at $T = 0$ plotted as a function of $2n_b$ for $J_o/2D = 0.05$, $|I_o|/2D = 0$ and various values of $E_o/2D$.

**Fig. 12.** The superconducting critical temperature $k_B T_c/2D$ of LP subsystem as a function of $E_o/2D$ plotted for $J_o/2D = 0.05$, $|I_o|/2D = 0$ and various values of $2n_b$. Solid and dashed lines mark second- and first order transitions, respectively, the tricritical points are given by circles.

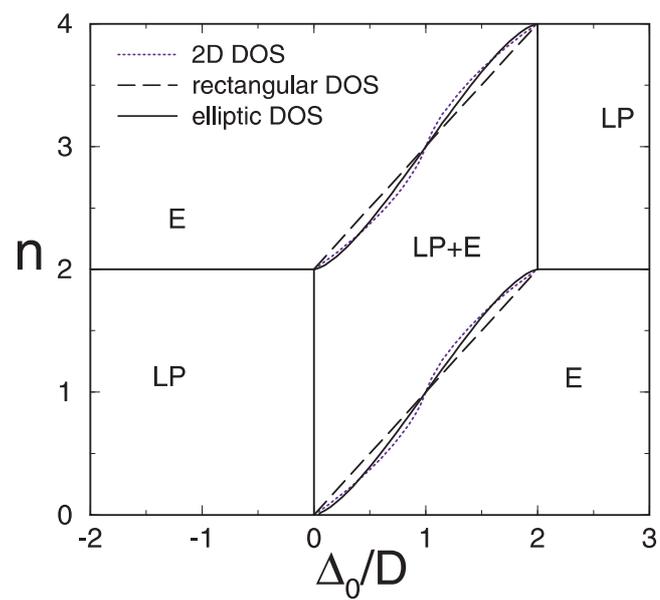

Fig. 2

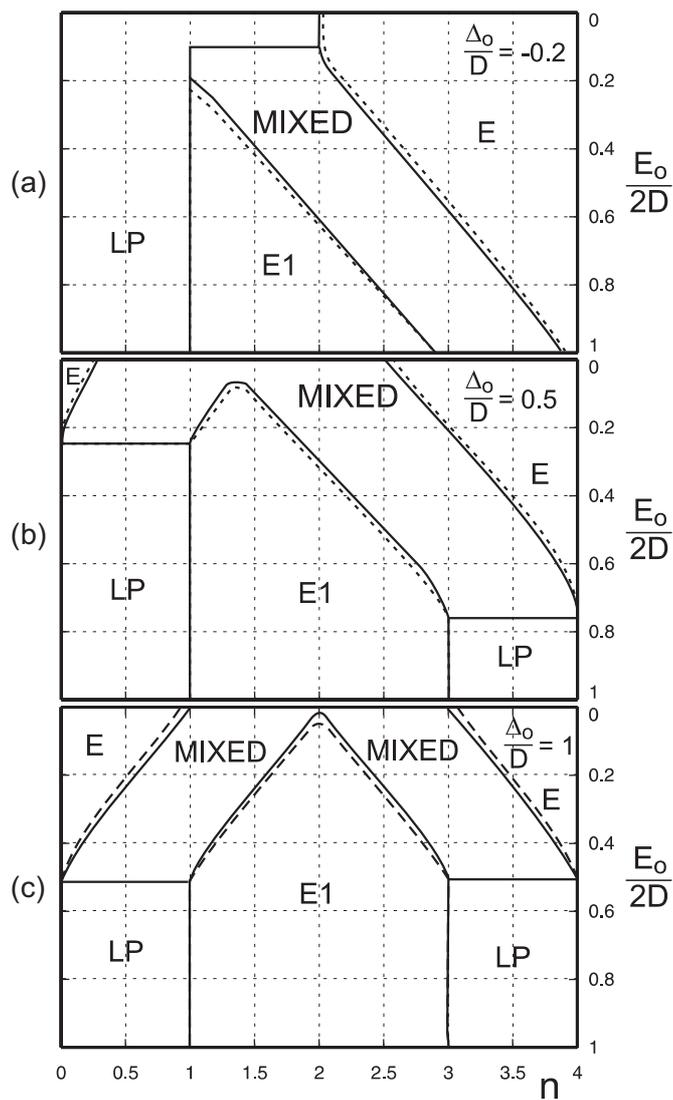

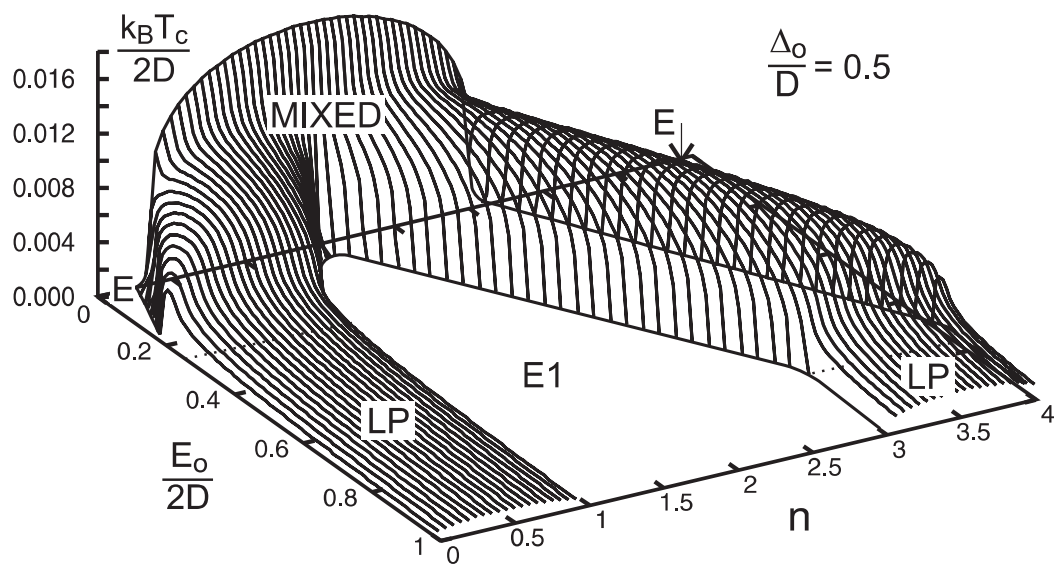

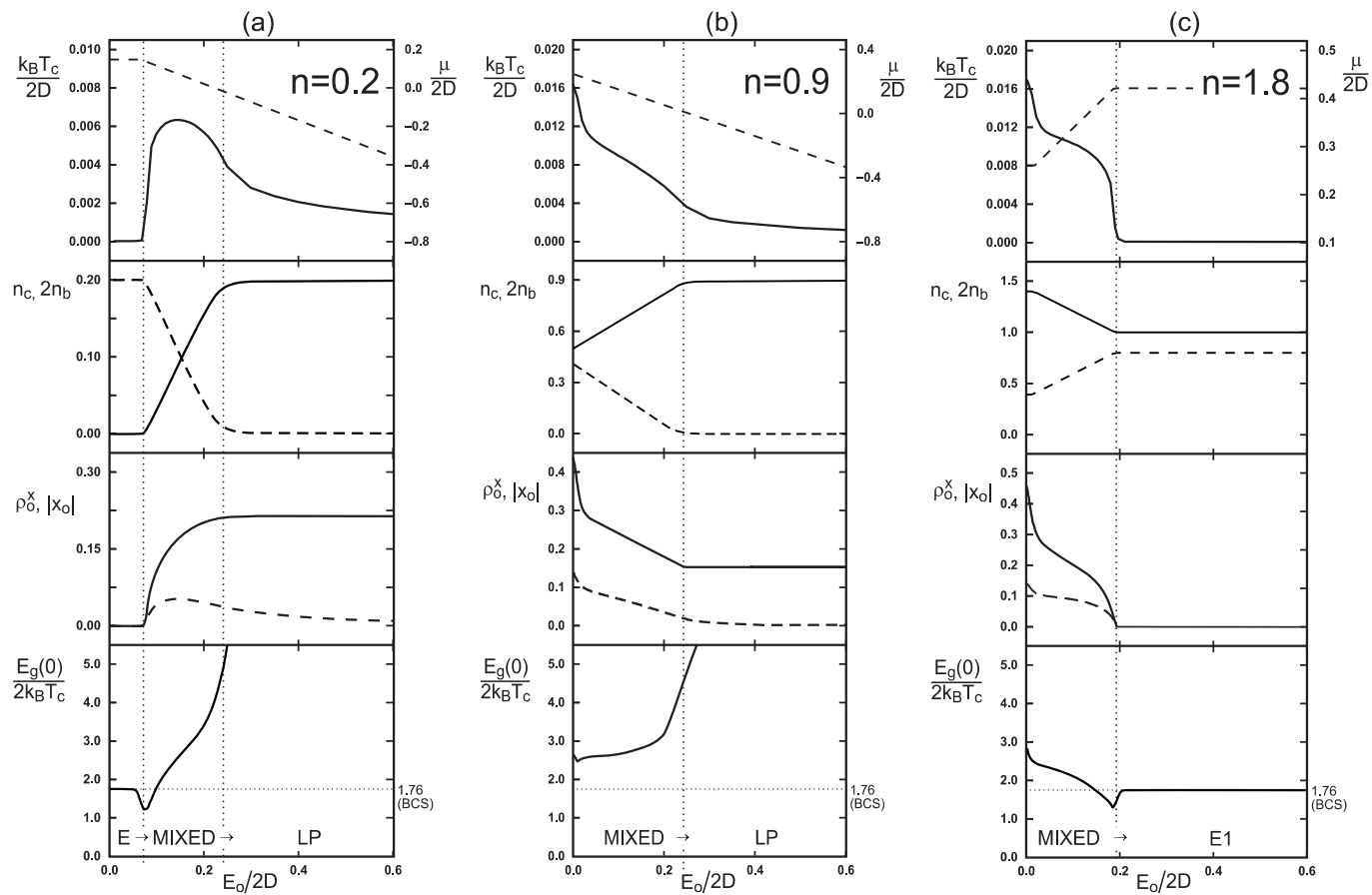

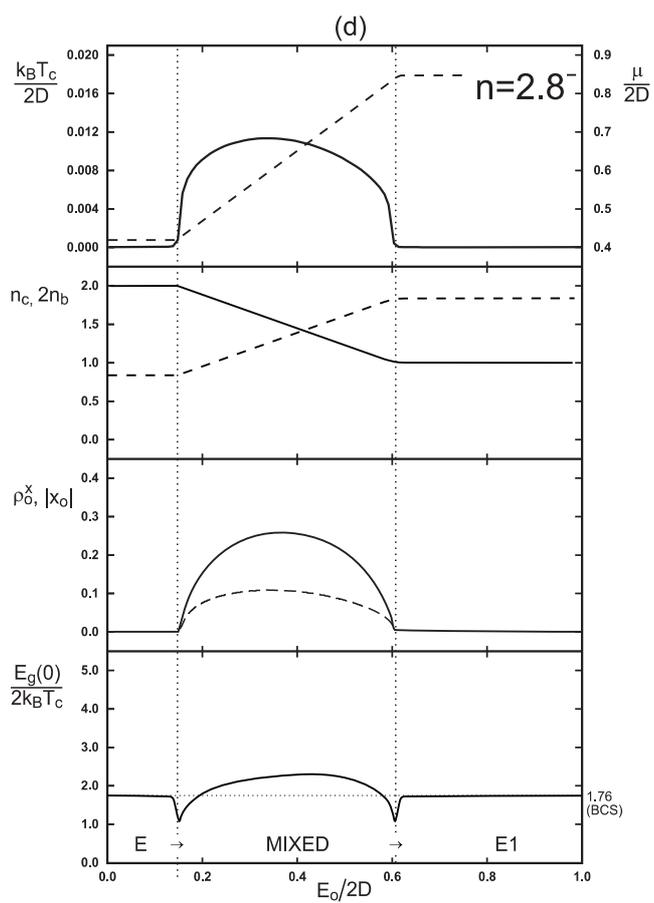
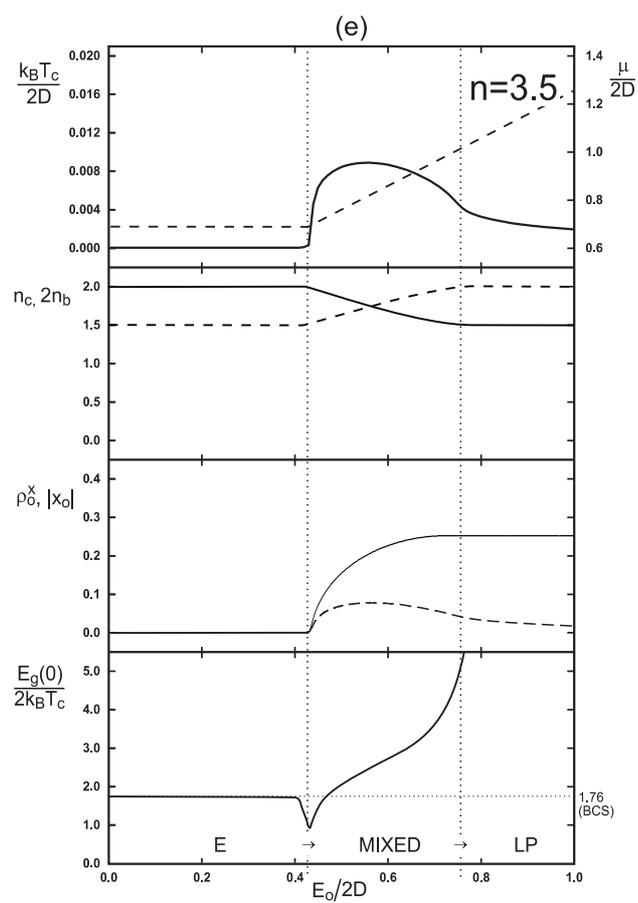

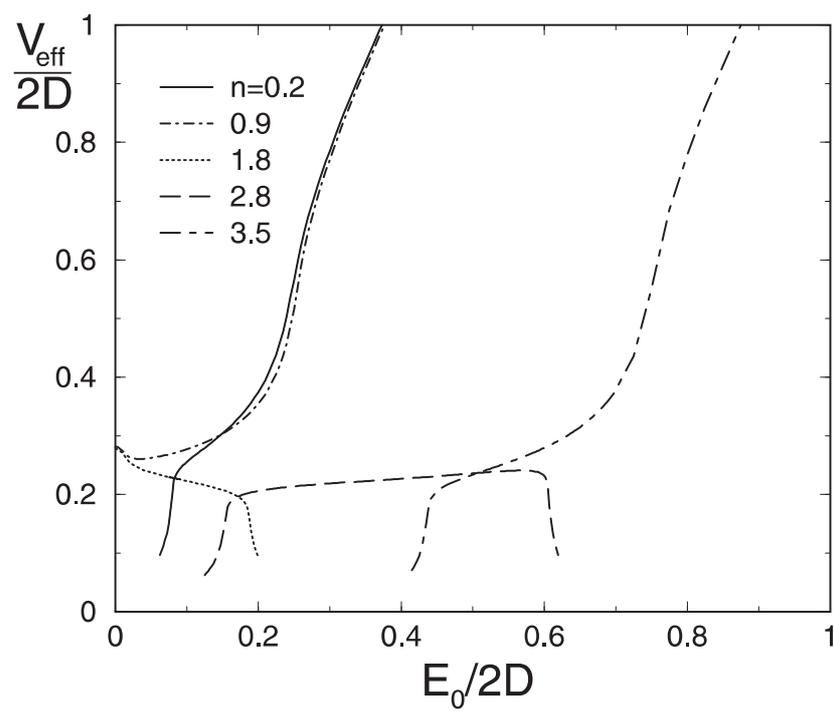

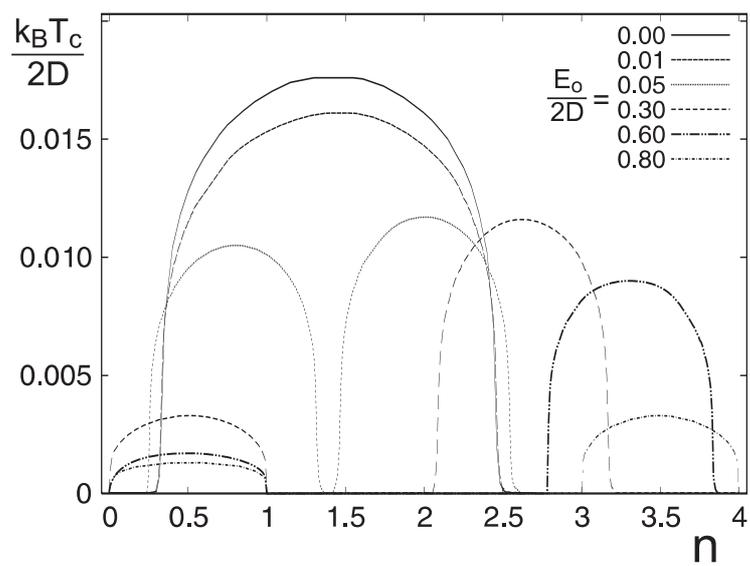

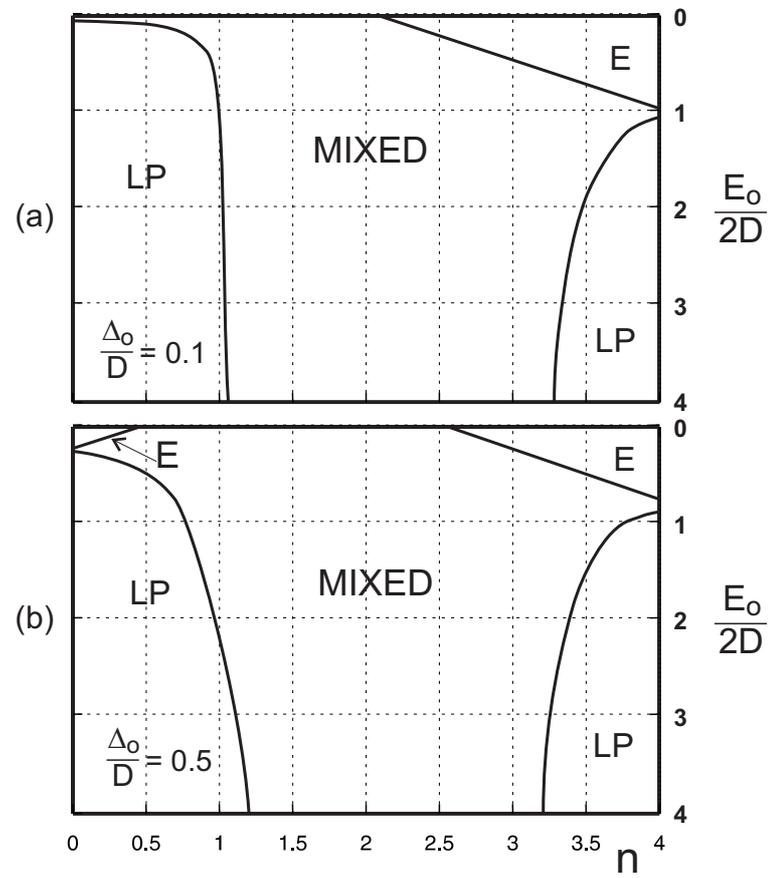

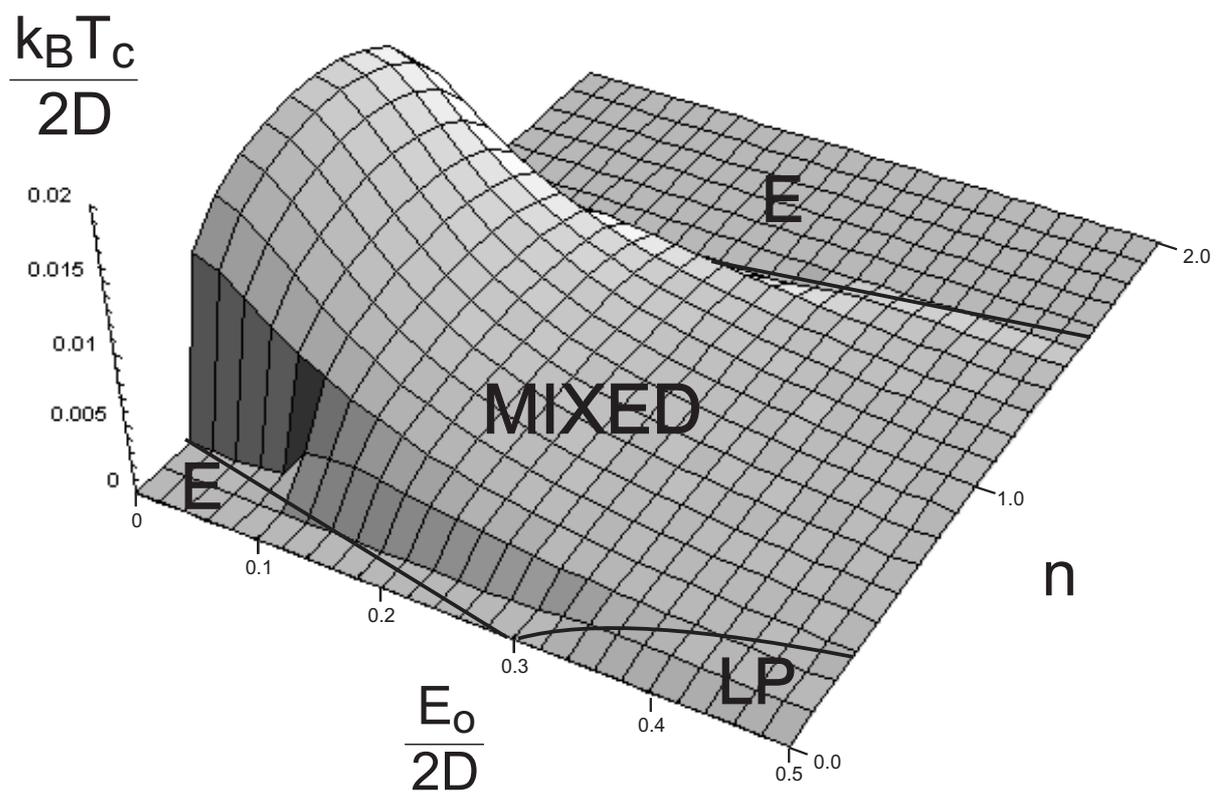

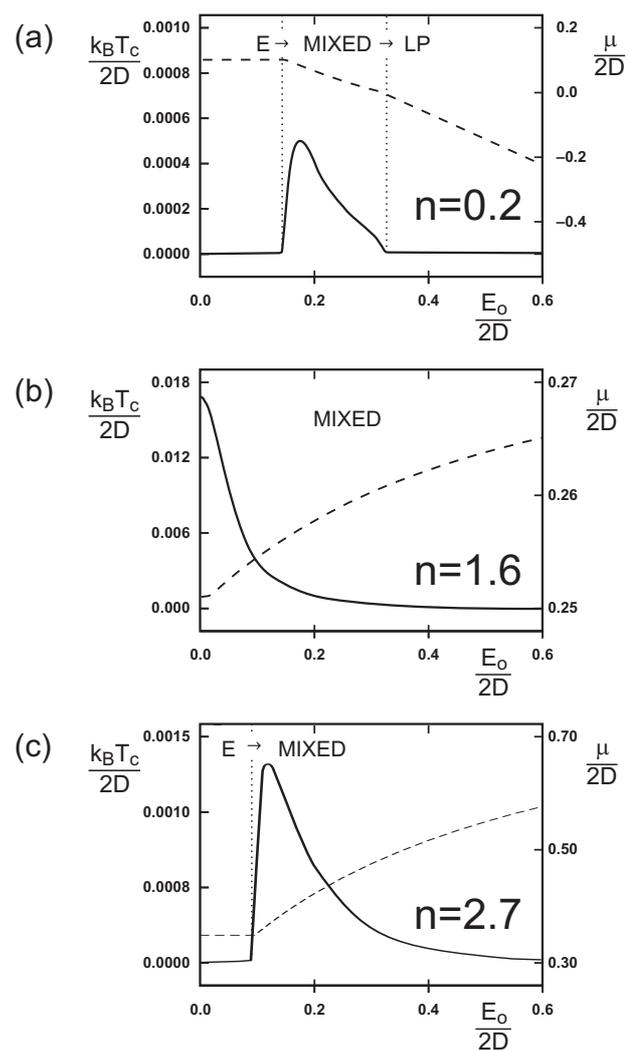

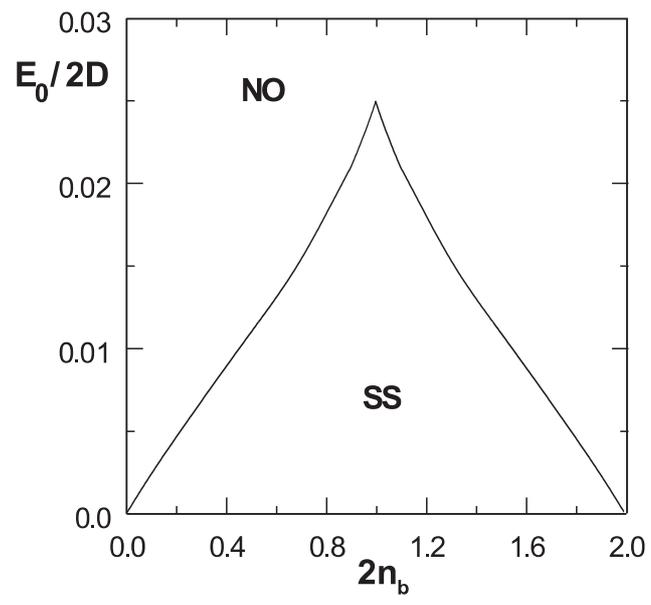

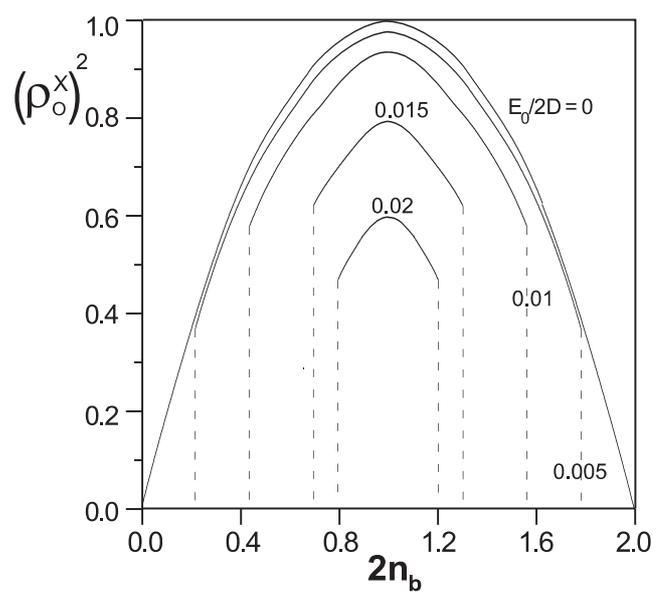

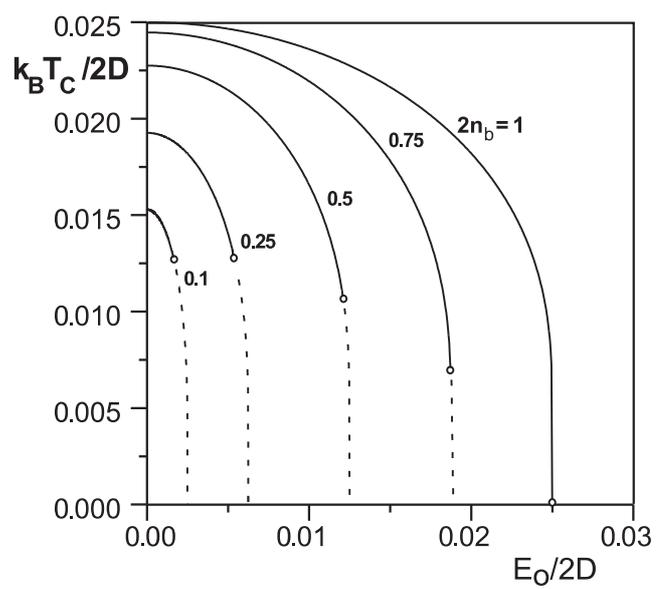